\documentclass[11pt,twoside]{article}
\usepackage{asp2014}

\aspSuppressVolSlug
\resetcounters

\begin{document}

\title{Rotation of White Dwarf Stars}
\author{Steven D. Kawaler$^1$
\affil{$^1$Department of Physics and Astronomy, Iowa State University, Ames, IA 50011 USA; \email{sdk@iastate.edu}}}

% This section is for ADS Processing.  There must be one line per author.
\paperauthor{Steven D. Kawaler}{sdk@iastate.edu}{ORCID_Or_Blank}{Iowa State University}{Physics and Astronomy}{Ames}{IA}{50011}{USA}

\begin{abstract}
I discuss and consider the status of observational determinations of the rotation velocities of white dwarf stars via asteroseismology and spectroscopy. While these observations have important implications on our understanding of the angular momentum evolution of stars in their late stages of evolution, more direct methods are sorely needed to disentangle ambiguities.
\end{abstract}

\section{White dwarf rotation - relics of main sequence angular momentum}
Rotation of white dwarfs provides an important boundary condition for the angular momentum distribution in stars from the main sequence through the asymptotic giant branch 
\citep{Kawa04, TayPin13, Cant14}. Despite decades of effort, our understanding of angular momentum evolution of ``normal'' single stars is incomplete. Complicating this effort is the difficulty of measuring rotation periods of stars. 
%For main-sequence stars, spectroscopic period determination requires sufficient spectral resolution to measure Doppler-broadening. 
Main-sequence stars with magnetic activity allow photometric period measurement
% (both in white light and with narrow-band photometry), 
but this requires precision photometry and sufficient temporal coverage to unambiguously determine a rotation period in the face of changing spot morphology. This is difficult from the ground, but data from {\it Kepler} has revolutionized our study of rotation of main sequence stars  \citep{McQu13, McQu14, Niel13, WalBas13}. On the main sequence, these data have enabled gyrochronological age determinations that use calibrated angular momentum loss formalisms \citep{WalBas13, Meib11}. 

Angular momentum transport between the core and envelope on the RGB and AGB should slow the rotation of the core, which would otherwise spin up to very short periods \citep{Kawa04}. Rotation beyond the main sequence has been examined using {\it Kepler} data by \citet{vonPin13} among others. Asteroseeismology has proven to be a viable tool for some RGB stars \citep{Beck12,Moss12}.  These studies suggest that significant angular momentum transport occurs between the shrinking core and the expanding envelope.  
% Beyond the RGB, photometric data on the rotation of stars include Kepler-enabled  measurement of sdB stars. Here again, asterosesimology plays a role, measuring the remnant surface rotation of sdB stars which have survived the core helium flash at the end of the RGB.  However, as leading ideas on the origin of sdB stars involve common-envelope binary evolution, it is not entirely clear that their rotation is representative of single star evolution \citep[though see][]{Pabl12}.
Measuring the rotation period of white dwarfs  provides useful constraints on the net angular momentum transport from the core to the envelope of stars during their post-main sequence evolution. As reviewed by \citet{Kawa04}, most reliable rotation periods from white dwarfs are from asteroseismology. Magnetic white dwarfs (usually with fields in excess of 10MG) also provide rotation periods through modulation of their spectra (and/or flux), but asteroseismology provides period determinations for ``normal'' white dwarfs.

% \section{What to expect for white dwarf rotation - what is fast and what is slow?}

Using observations of main sequence rotation periods as a function of stellar mass, and some simple limiting cases of angular momentum transport within stars, we can place interesting limits on how fast white dwarfs {\it could} spin. The simplest limiting case is for a core whose angular momentum remains uncoupled to the envelope (except when linked by convection). This kind of analysis has been presented by \citet{Kawa04}, with more sophisticated scenarios using evolutionary models adding more details \citep[i.e.][]{TayPin13}. The basic results are summarized in Table 1.  Low mass stars must produce white dwarfs that have rotation periods longer than 5 hours. More massive stars produce white dwarfs that could have rotation periods as short as several minutes.  In reality, though, any angular momentum coupling between the core and envelope on the giant branch will produce more slowly rotating cores.  However, if the coupling is sufficiently strong,  the remnant white dwarf could have a rotation period as long as years \citep{Kawa04, TayPin13}.

\begin{table}[!ht]
\caption{Lower limits to rotation period $P$ in the cores of stars.  $J$ denotes total angular momentum for the star, 
HB is an abbreviation for Horizontal Branch.}
\begin{tabular}{cccc}

\noalign{\smallskip}
\tableline
Evolutionary Stage & $M < 1.2 M_{\odot}$  & $1.2 M_{\odot} <M< 2.3 M_{\odot}$  & $M > 2.3 M_{\odot}$ \\
\noalign{\smallskip}
\tableline
Main Sequence &  $dJ/dt$ on M.S. & no $dJ/dt$ & no $dJ/dt$ \\
           &  slow start,                       & fast start,   & fast start, \\
           & $P_{\rm rot} \approx 20$d   & $P_{\rm rot} \approx 20$h & $P_{\rm rot} \approx 20$h\\
\tableline
RGB to He ign. & $\Delta M, \Delta J$ at RGB tip  & $\Delta M, \Delta J$ at RGB tip & no $M$, $J$ loss\\
                               &  $P_{\rm rot} \approx 5$h & $P_{\rm rot} \approx 0.7$ h&  $P_{\rm rot} \approx 0.7 $ h \\
\tableline
Core He burning  & degen. flash & degen. flash & non-degen \\
                             & HB & HB/clump & clump \\
                          & $P_{\rm rot} \approx 50$ h  & $P_{\rm rot} \approx 7$h  & $P_{\rm rot} \approx$ 0.7 h \\
\tableline
AGB / post-AGB    & env. $J$, $M$ loss &  env. $J$, $M$ loss &  env. $J$, $M$ loss \\
                              & $P_{\rm rot} \approx 5$ h       &  $P_{\rm rot} \approx 0.7$ h  & $P_{\rm rot} \approx 0.07$ h\\
\tableline
WD mass  &  $M_{\rm WD} < 0.53M_{\odot}$  & $0.53M_{\odot} < M_{\rm WD}$ & $M_{\rm WD} > 0.65 M_{\odot}$\\
                               &                                                 & $M_{\rm WD} < 0.65M_{\odot}$  & \\
\tableline
\end{tabular}
\end{table}
                            
\section{Measuring white dwarf rotation periods}

For all but the shortest conceivable periods, rotational broadening of white dwarf spectral lines  is dwarfed by the broadened lines in these high-gravity stars. The bulk of observed white dwarf rotation periods have been determined via photometry and spectroscopy of magnetic white dwarfs, and through asteroseismology.

Magnetic white dwarfs comprise approximately 20 percent of know white dwarf stars \citep{Kawk07, Kepl13}.
%, with the incidence increasing at lower fields \citep{Lieb03}. 
Given the intense surface gravity of white dwarfs, the surface field needs to be of order a MG or greater to influence the spectrum sufficiently for a field strength determination via spectroscopy; spectropolarimetry can reveal fields in the tens-of-kG range \citep{Kawk07}. Simple scaling and stellar population arguments \citep[i.e.][]{WicFer00} suggest that these fields are difficult to generate through dynamo activity, and are likely to be remnant fields from earlier evolution as magnetic A stars on the main sequence. Manifestations of these large magnetic fields include time-dependent line broadening and polarization, and broader photometric effects (i.e. from starspots).  For normal white dwarfs, near-surface dynamo activity in a convection zone could produce magnetic spots (hot or cold), which could reveal rotation through photometric variation. 

\citet{Brink13} observed 30 isolated magnetic white dwarfs.  They found photometric variability in nine stars with fields below 10MG, including two well below 1MG. These $\approx$1\% variations were uncorrelated in period or amplitude with  field strength or $T_{\rm eff}$. Periods ranged from less than an hour to 4 days, with two stars showing much longer periods. \citet{HowHol11} found a 0.26 day period in Kepler photometry of a white dwarf with a 0.3MG field.  Rotation periods for magnetic white dwarfs, from \citet{Kawk07} and later sources, are given in Table 2.

\begin{table}[!ht]
\caption{Measured rotation periods of single magnetic white dwarfs}

\begin{tabular}{lccccc} 
\tableline
Star & $P_{\rm rot}$ [h] & Type & $M$ [$M_{\odot}$] & $B$ [MG] $T_{\rm eff}$ \\
\tableline
PG 1015 & 1.6 & DA & 0.6 & 120 & 14,000 \\
HE 1211 & 2.0 & DB & 0.6 & 50 & 12,000 \\
Feige 7 & 2.2 & DAB & 0.6 & 35 & 20,000 \\
HE 1045 & 2.7 & DA & 0.6 & 16 & 10,000 \\
PG 1031 & 3.4 & DA & 0.6 & 600 & 15,000 \\
PG 1312 & 5.4 & DA & 0.6 & 10 & 20,000 \\
BPM 25114 & 68 & DA & 0.6 & 36 & 20,000 \\
KUV 813-14 & 429 & DA & 0.6 & 45 & 11,000 \\
SDSS J000555.90 & 51 & DQ & 0.6 & 1.47 & 19,400 \\
GD 356 & 2.0 & DB & 0.67 & 13 & 7,510 \\
G99-37 & 4.1 & DQ & 0.67 & 10 & 6,070 \\
G99-47 & 1.0 & DA & 0.71 & 29 & 5,790 \\
BOKS 53856 & 6.14 & DA & 0.68 & 0.35 & 32,500 \\
G92-40 & 35 & DA & 0.74 & 0.07 & 7,920 \\
G195-19 & 32 & DB & 0.75 & 100 & 7,160 \\
EUVE J0317-855 & 0.2 & DA & 1.35 & 300 & 33,000 \\
\tableline
\end{tabular}
\end{table}

Another sample that has yielded rotation period measurements consists of the nonradially pulsating white dwarfs (summarized in Table 3). Briefly, nonradial oscillation frequencies in an axisymmetric star are degenerate for various values of the azimuthal quantum number $m$ for modes of the same degree $l$ and order $n$. A break in that azimuthal symmetry (i.e. rotation) will cause a separation in frequency between modes of differing $m$ by an amount equal to $m \Omega (1-C)$ where $\Omega$ is the rotation frequency and  $C$ depends on the internal structure of the star. So, a multiplet emerges with a splitting in frequency  that is proportional to  $\Omega$. For white dwarfs, $C \approx (l[l+1])^{-1}$.  Thus in practice, rotationally split $l=1$ modes become triplets, with a spacing equal to $\Omega/2$. It is this signature that reveals the white dwarf rotation periods reviewed in \citet{Kawa04} and \citet{FonBra08}.  For rotation periods shorter than several hours, the regular period spacings (separated by successive values of $n$) can overlap the rotationally split multiplets.  Figure 1 (left panel) shows this  short rotation period confusion limit for the three most common types of pulsating white dwarfs.  

\begin{table}[!ht]
\caption{Rotation periods of white dwarfs as determined via asteroseismology}

\begin{tabular}{lcccc}  % l = left, c = centered
\tableline
\noalign{\smallskip}
Star & $P_{\rm rot}$ [h] & $v_{\rm rot}$ [km/s] & Type & $M$ [$M_{\odot}$] \\
\noalign{\smallskip}
\tableline
PG 0122   & 37 & 0.66 & GW Vir & 0.56 \\
NGC 1501& 28 & 0.87 & GW Vir & 0.56 \\
PG 1707   & 16 & 1.53 & GW Vir & 0.56 \\
RX J2117 & 28 & 0.87 & GW VIr & 0.57 \\
PG 1159 & 33 & 0.74 & GW Vir & 0.60 \\
PG 2131 & 5 & 4.89 & GW Vir & 0.60 \\
EC 20058 & 2 & 8.73 & DBV & 0.54 \\
KIC 8626021 & 41 & 0.43 & DBV & 0.56 \\
GD 358 & 29 & 0.60 & DBV & 0.61 \\
HL Tau 76 & 53 & 0.33 & C-ZZ Ceti & 0.55 \\
KIC 11911480 & 84 & 0.21 & H-ZZ Ceti & 0.57 \\
R548 & 37 & 0.47 & H-ZZ Ceti & 0.60 \\
HS0507 & 41 & 0.43 & C-ZZ Ceti & 0.6 \\
G29-38 & 32 & 0.55 & C-ZZ Ceti & 0.6 \\
GD 165 & 50 & 0.35 & H-ZZ Ceti & 0.63 \\
KUV11370+4222 & 5.56 & 3.14 & C-ZZ Ceti & 0.63 \\
G185-32 & 15 & 1.16 & H-ZZ Ceti & 0.64 \\
GD 154 & 55 & 0.32 & C-ZZ Ceti & 0.70 \\
L19-2 & 13 & 1.34 & H-ZZ Ceti & 0.71 \\
EC14012-1446 & 14.4 & 1.21 & CH-ZZ Ceti & 0.71 \\
G226-29 & 9 & 1.94 & H-ZZ Ceti & 0.78 \\
J1612+0830 & 0.93 & 18.77 & ZZ Ceti & 0.8 \\
J1916+3936 & 18.8 & 0.93 & ZZ Ceti & 0.82 \\
J1711+6541 & 16.4 & 1.06 & ZZ Ceti & 1.00 \\
\tableline % Sometimes you just need a line between table rows
\end{tabular}
\end{table}

Asteroseismic rotation periods are weighted averages of the internal rotation profile with depth. With a  set of  modes that sample different parts of the stellar interior one can, in principle, measure differential rotation in stars.  
%This is how \citet{Beck12} and \citet{Moss12} measured differential rotation in RGB stars. 
Early attempts to test differential rotation in white dwarfs \citep[i.e.][]{Kawa99} were inconclusive; recent attempts remain ambiguous \citep{Charp09, Cors11}.  

\section{Observed rotation periods and their implications}

The right panel in Figure 1 is a  histogram of the periods measured via asteroseismology and photometry of magnetic white dwarfs, we see a distinct peak at around 1-2 days, with a median value of 28 hours. Magnetic white dwarfs cover a broader range, with most having periods less than one day.  The distribution cannot be considered to represent the underlying distribution of white dwarf rotation periods. 

\articlefiguretwo{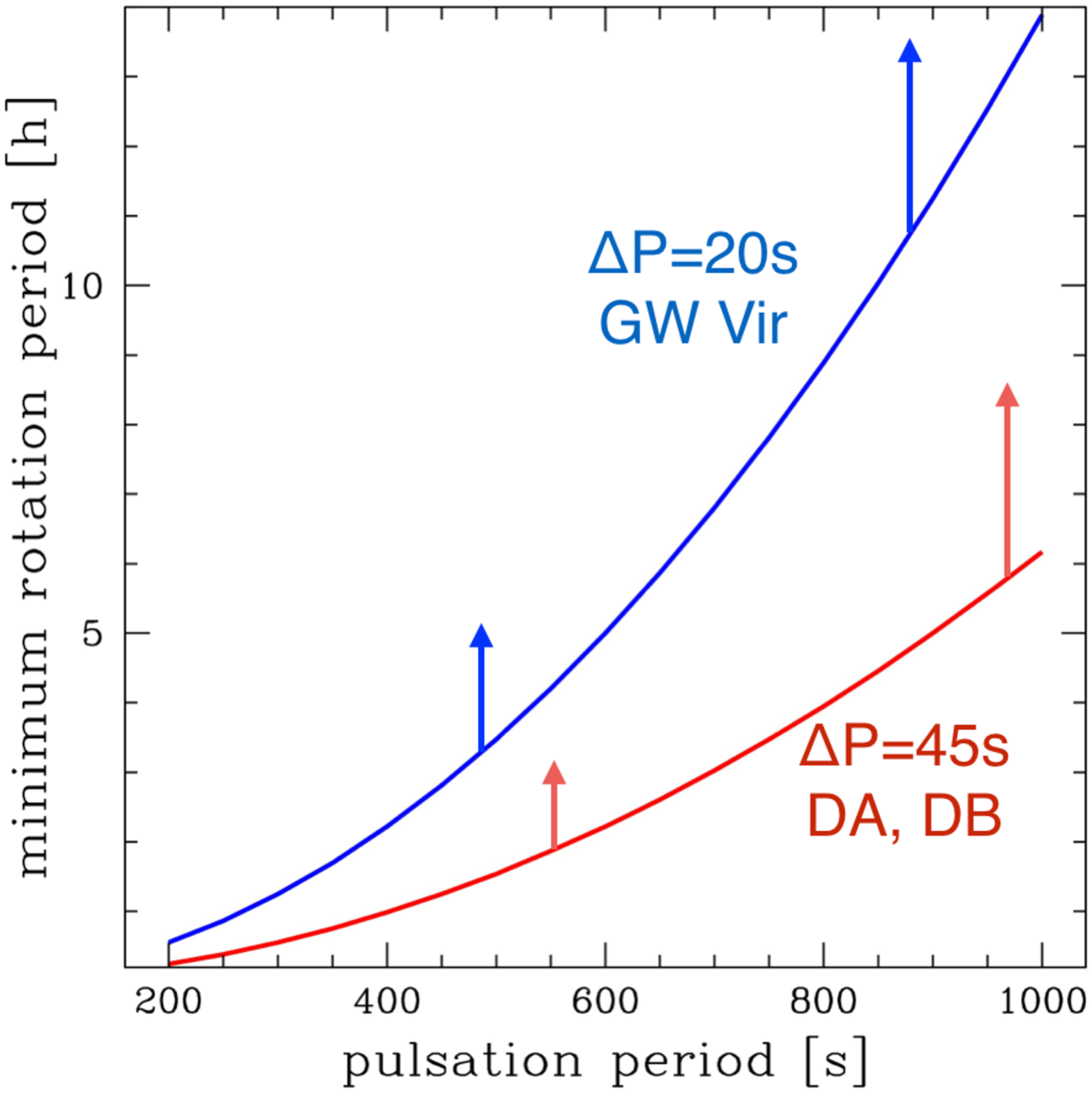}{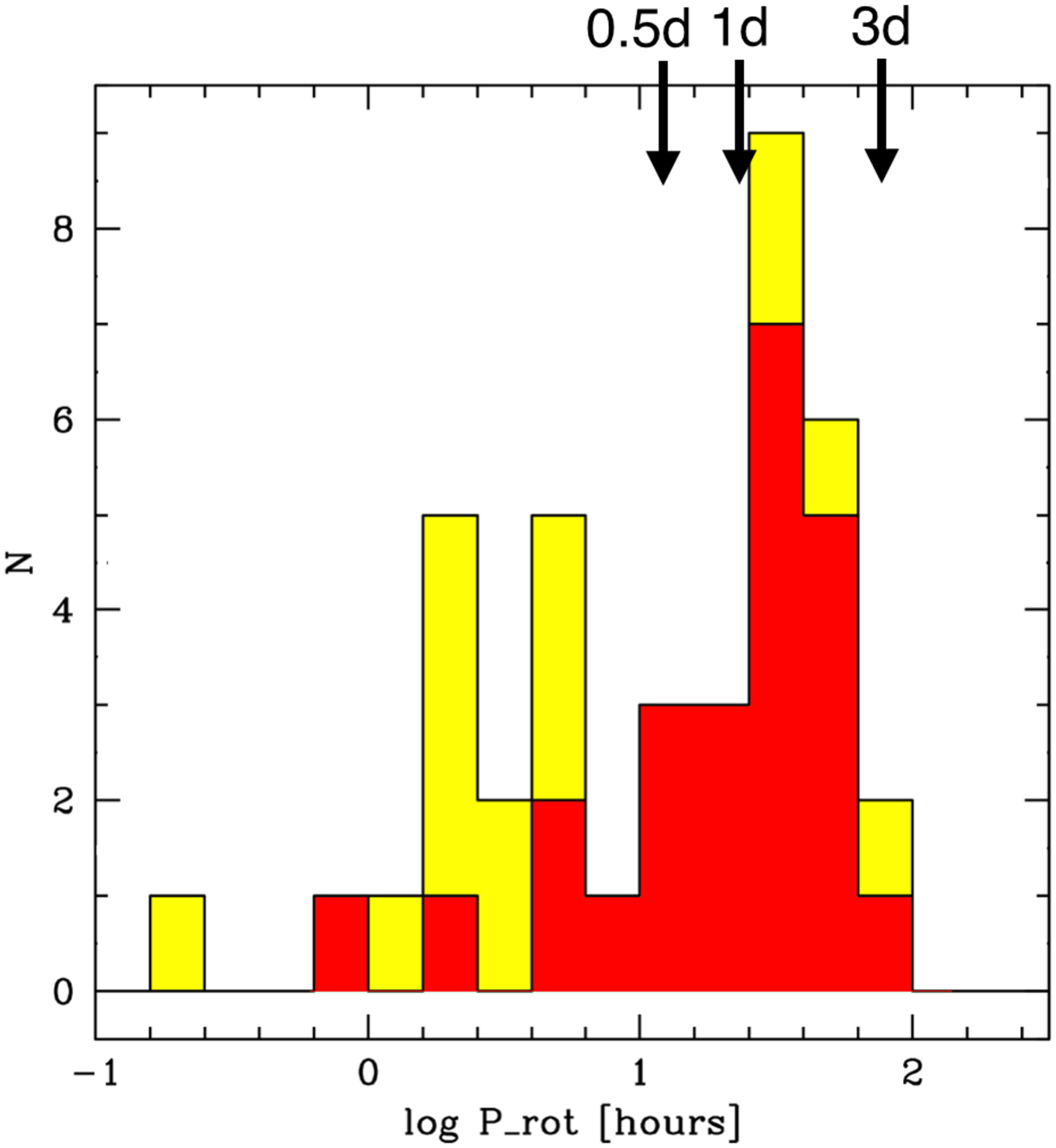}{fig12}{ \emph{Left:} Minimum rotation period that can be unambiguously determined via asteroseismology
% without overlap between $l$=1 mode triplets and the natural period spacing of $g$ modes for white dwarf stars.  
\emph{Right:} Histogram of rotation periods.  Yellow (white) are magnetic white dwarfs, while red (dark) denotes asteroseismic periods.}

While selection effects in this sample are horrendous, this at least demonstrates that white dwarfs {\em for which periods can be determined} rotate ``slowly'' in general.  But the fact that the median rotation period is as short as a day or two means that angular momentum transport between the core and envelope is insufficient to enforce solid-body rotation on the RGB or the AGB.  \citet{Cant14} explore angular momentum transport in evolutionary models; such efforts may be able to accommodate this residual core rotation, but a larger sample of rotation rates for white dwarfs with better--understood systematics is needed.  

Asteroseismology holds significant promise for determining a consistent white dwarf rotation period distribution -- all white dwarfs are expected to pulsate when the reach the appropriate effective temperature range.  The {\it Kepler} mission had the potential to reveal rotational modulation in white dwarfs.
%but normal white dwarfs were relatively scarce in the original Kepler field. 
Several were included in the initial {\it Kepler} survey phase \citep{Oste10};
Recently, \citet{Maoz14} reported  that roughly half of those targets (7 out of 14) show photometric variations at the 0.1 to 1 ppt. The {\it K2} mission may observe a large sample new (and known) pulsating white dwarfs.  This sample could be indispensable for validating asteroseismic rotation periods and for determining the underlying distribution of white dwarf rotation periods.

%\clearpage % To force this stuff to happen by this point in the text, otherwise these will probably end up after the references.

%\acknowledgements The ASP would like to the thank the dedicated researchers who are publishing with the ASP.  Keep this text on the same line as the \verb"\acknowledgements" command because it makes things a lot easier.

% For non-BibTex:

\end{document}